\documentclass{article}

\usepackage{arxiv}

\usepackage[utf8]{inputenc} 
\usepackage[T1]{fontenc}    
\usepackage{hyperref}       
\usepackage{url}            
\usepackage{booktabs}       
\usepackage{amsfonts}       
\usepackage{nicefrac}       
\usepackage{microtype}      
\usepackage{lipsum}		
\usepackage{graphicx}
\usepackage{natbib}
\usepackage{doi}
\usepackage{caption}
\usepackage{subcaption}
\usepackage{algorithm}
\usepackage{algorithmic}
\usepackage{amsmath}

\title{A Digital Compositing Approach to obtain Animated Chinese Still-life Paintings with Global Effects}

\date{} 					

\author{ 
 	 Sitong Deng\\
Department of Visualization\\ 
Texas A\&M University, College Station, TX, 77831\\
	\texttt{sitongdeng@tamu.edu} \\
    \And
    \href{https://orcid.org/0000-0003-3618-4166}{\includegraphics[scale=0.06]{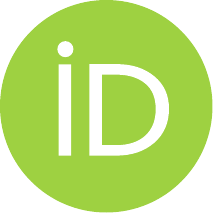}\hspace{1mm}Ergun Akleman}\thanks{Joint with Computer Science and Engineering Department.} \\
	Visual Computing \& Computational Media,\\ Texas A\&M University, College Station, TX, 77831\\
	\texttt{ergun@tamu.edu} \\
}

\renewcommand{\shorttitle}{A Dynamic Chinese still-life Painting}

\hypersetup{
pdftitle={A template for the arxiv style},
pdfsubject={q-bio.NC, q-bio.QM},
pdfauthor={David S.~Hippocampus, Elias D.~Striatum},
pdfkeywords={First keyword, Second keyword, More},
}

\begin{document}
\maketitle

\begin{abstract}
In this work, we present a method for turning Chinese still-life paintings with global illumination effects into dynamic paintings with moving lights. Our goal is to preserve the original look and feel of still-life paintings with moving lights and objects. We have developed a deceptively simple method that can be computed as a composite of two animated texture images using an animated rendering. The compositing process can be implemented directly in an animation system such as AfterEffect, which allows for the basic compositing operation over animations. It is also possible to control the colors by changing the material colors in animated rendering. We have provided a proof-of-concept based on an original digital Still-Life painting that is in realist Chinese style. This approach can be used to turn almost any still-life painting into a dynamic painting. 
\end{abstract}

\section{Introduction and Motivation} 

\begin{figure}[htb]
        \includegraphics[width=1.0\textwidth]{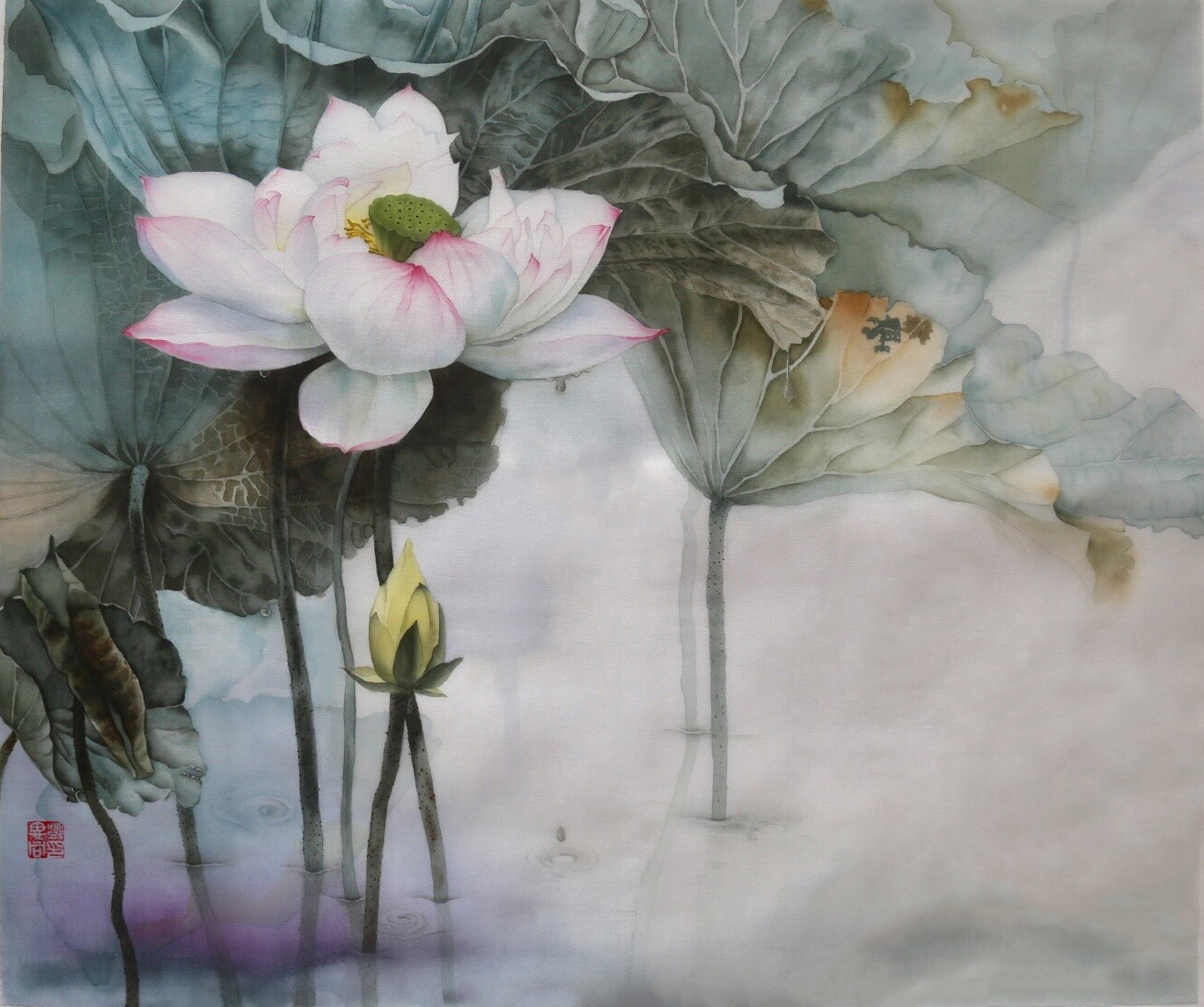}
        \caption{A Chinese Painting by Sitong Deng. Our goal is to create dynamic versions of paintings such as this one just by using basic compositing operators. This approach can allow artists to have direct control to emulate the look and feel of their own paintings. }
        \label{fig_Original_Painting}
\end{figure}

\begin{figure}[htb]
         \begin{subfigure}[t]{0.48\textwidth}
        \includegraphics[width=1.0\textwidth]{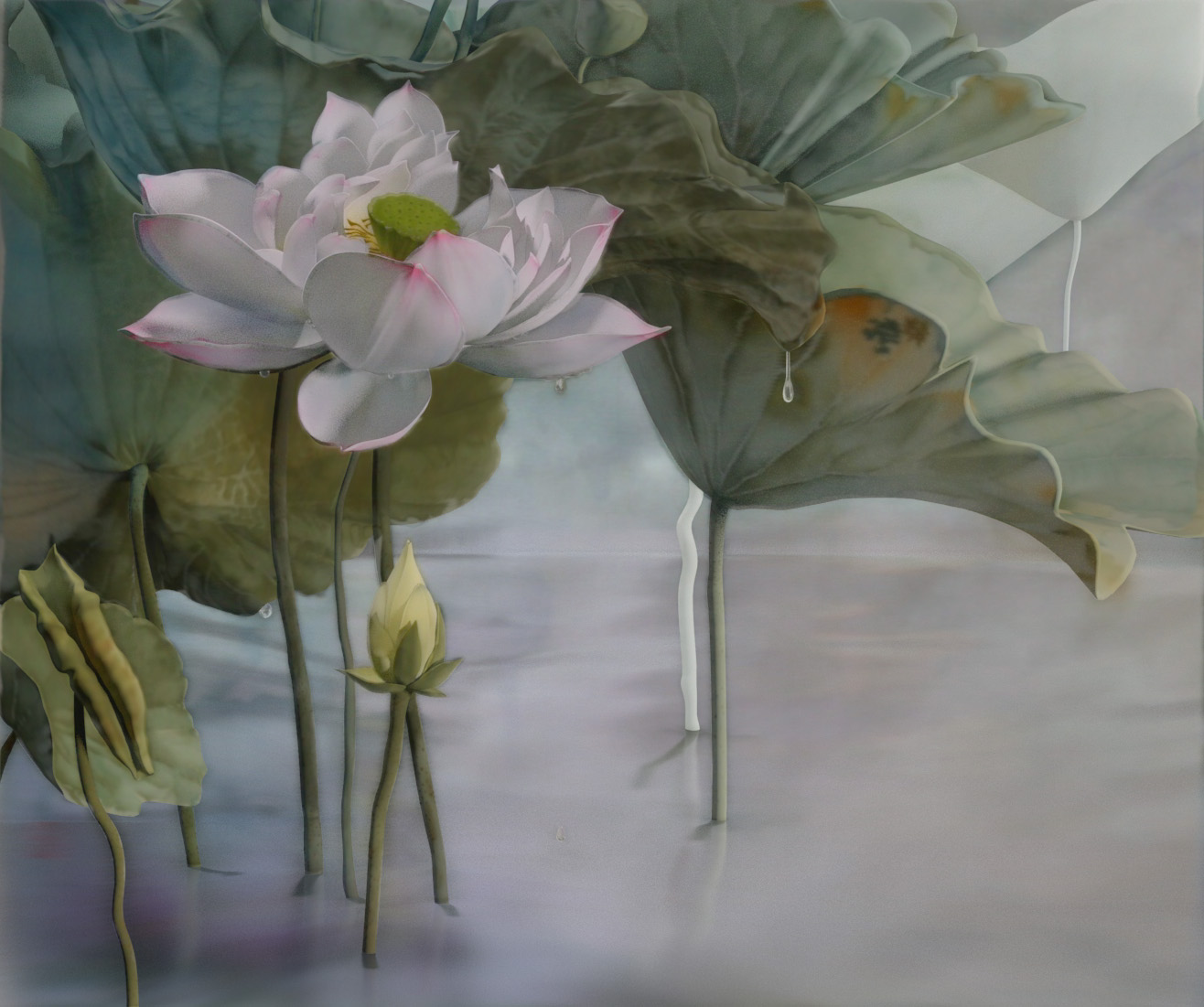}
        \caption{Frame 26 of our animation.}
        \label{fig_frame26}
    \end{subfigure}
    \hfill   
         \begin{subfigure}[t]{0.48\textwidth}
        \includegraphics[width=1.0\textwidth]{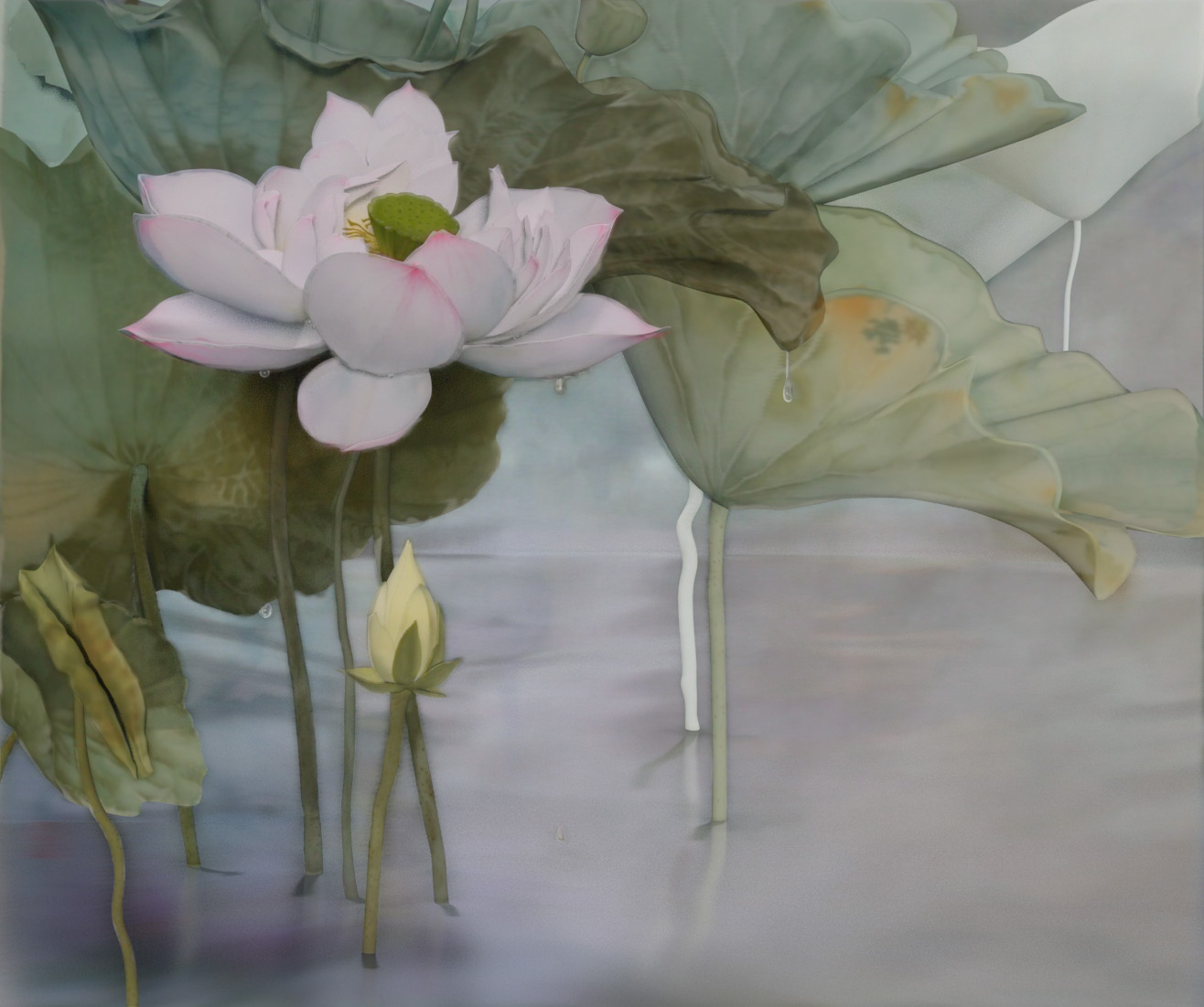}
        \caption{Frame 56 of our animation.}
        \label{fig_frame56}
    \end{subfigure}
    \caption{\it An example of emulating a still-life painting with dynamically changing shadows, specular highlights, and caustics using a moving area light source. Our method allows direct artistic control to emulate look-and-feel of the original painting. }
\label{fig_teaser}
\end{figure}

\begin{figure}[htb]
        \begin{subfigure}[t]{0.48\textwidth}
        \includegraphics[width=1.0\textwidth]{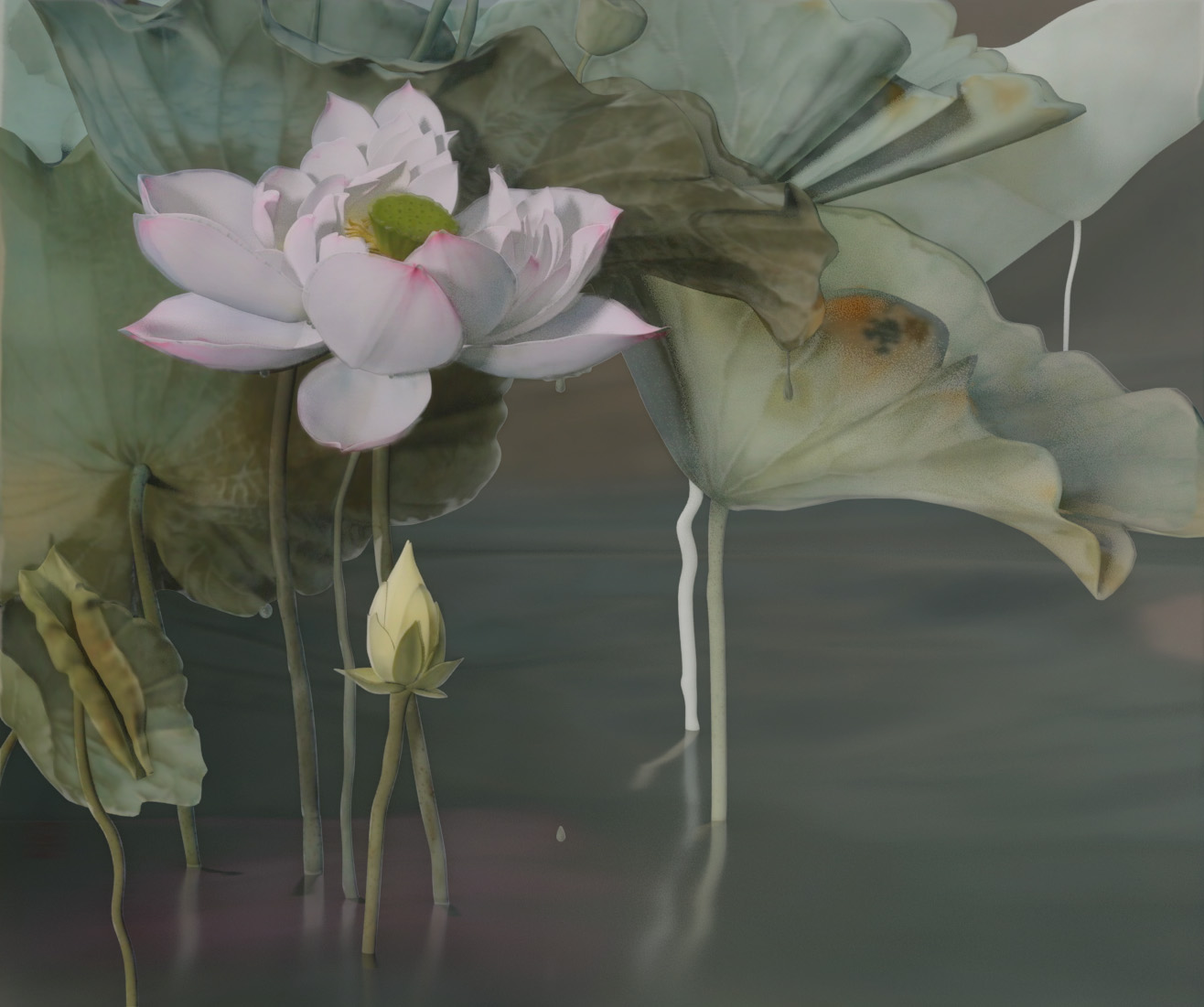}
        \caption{Frame 72 of our animation.}
        \label{fig_frame72}
    \end{subfigure}
    \hfill 
    \begin{subfigure}[t]{0.48\textwidth}
        \includegraphics[width=1.0\textwidth]{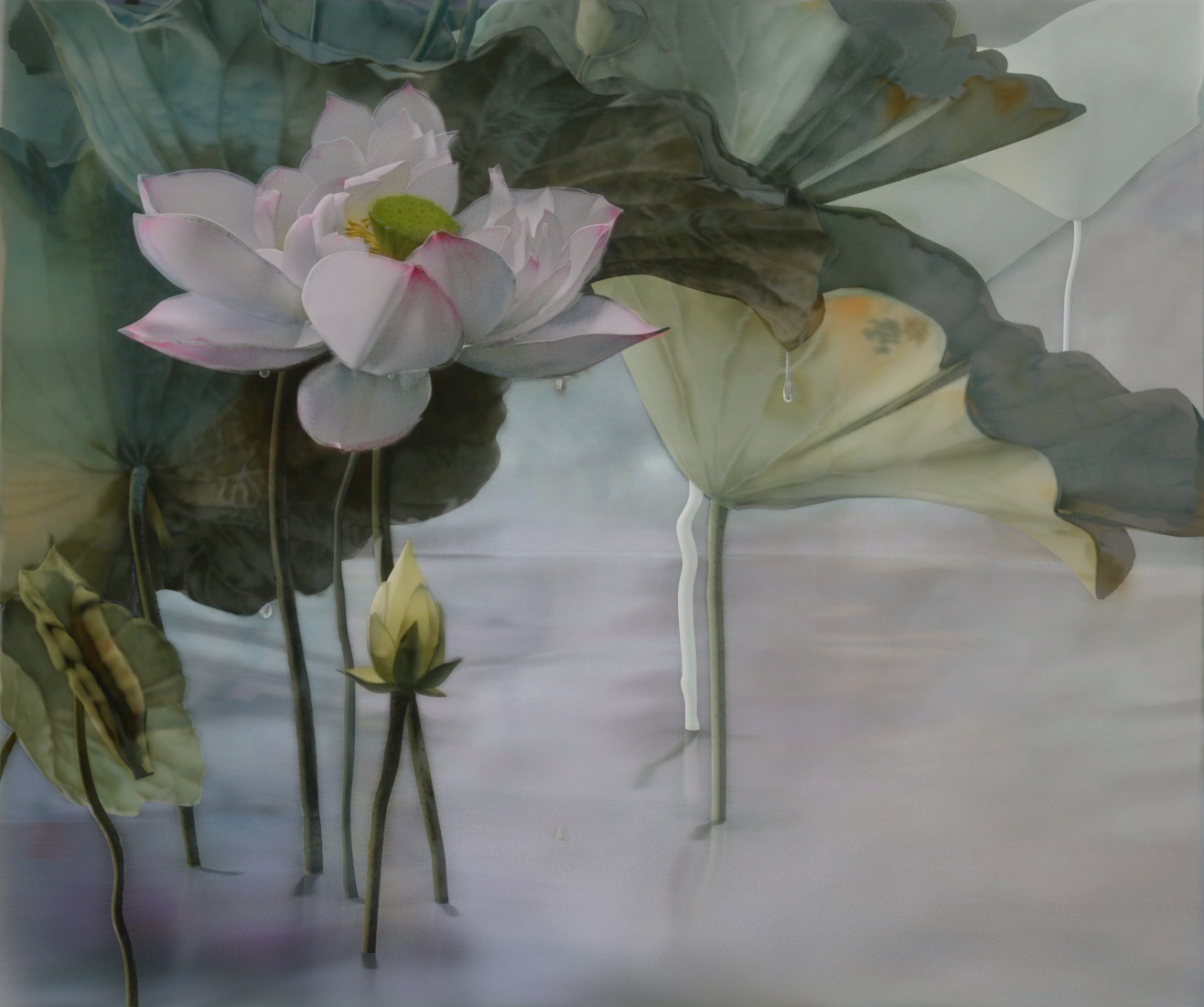}
        \caption{Frame 119 of our animation.}
        \label{fig_frame119}
    \end{subfigure}
    \hfill 
\caption{\it More examples of animated frames. }
\label{fig_teaser2}
\end{figure}

Global illumination effects cannot be created only by using local shader functions. The usual examples of global illumination effects are reflections, refractions, and caustics. We also include shadows, since they cannot be computed only by local shader functions. In all of these cases, the interrelations between objects play an important role. Global illumination effects, except shadows, are relatively rare in most of the subjects in paintings. Refractions are usually seen in painterly subjects, such as still life \cite{subramanian2020painterly}. Reflections are relatively more common. They exist in landscape paintings that include large bodies of water, such as lakes or seas \cite{zhao2015}. Chinese painting is not an exception. Most subjects in traditional Chinese painting include trees, birds, and mountains \cite{li2013rendering, wang2014stylization, li2023rendering}, but we rarely see paintings that include large bodies of water \cite{liu2015}. In this work, our goal is to create a dynamic version of a painting of water lilies in the Chinese painting style by Sitong Deng (see Figure~\ref{fig_Original_Painting}). As shown in the figure, this particular painting includes a pond of still water that reflects water lilies. The goal is to create an animated version of this painting that can provide subtle motion of water, water lilies, and the position of the sun. Figures~\ref{fig_teaser}. and~\ref{fig_teaser2} show frames of an animation that is created by using our compositing method. 

The main problem with the inclusion of global illumination effects in rendering comes directly from its definition. We need to compute interreflections between the objects to include global illuminations such as reflections and refractions, which requires a more powerful rendering approach such as ray tracing \cite{glassner1989introduction, cook1984distributed, shirley2008realistic}, photon mapping \cite{jensen2001realistic}, path tracing \cite{keller2015path, christensen2018renderman}, or radiosity \cite{cohen1993radiosity}.
For the creation of expressive and stylistic look-and-feel, these methods are not directly useful since they were developed to obtain realistic visuals, which are usually called "photo-realism".

During the last decade, our research group has developed a large number of methods to create dynamic paintings with global illumination effects using a variety of barycentric shading methods \cite{eisinger2012,wang2014a,akleman2015art, zhao2015, liu2015, xiong2016,du2016,du2017,castaneda2017}. However, all of these methods require an implementation at the shader level. Although basic shader operations are simple, this requirement makes the methods harder to implement for most naive users who are not familiar with shading programming. For users who are only familiar with commercial software such as Maya and AfterEffect, \textbf{there is a need for a simple methodology to create dynamic paintings with global illumination effects}. In this work, we have developed such a process by moving barycentric shading into the compositing stage. An important part of this process is that it is deceptively simple to use and provides art-directable control to users directly at the compositing level.

The main advantage of the new method is that it is independent of the type of rendering method used. Any rendering method from photon mapping to path tracing from ray tracing to radiosity can be used, as long as the method supports the desired global illumination. In this particular case, we want to have shadows and reflections. Since radiosity does not provide mirror reflections, path tracing, and photon mapping are overkill just to obtain reflections and shadows ray tracing seems to be sufficient. On the other hand, since our approach is agnostic to the rendering method, more powerful rendering approaches such as path tracing or photon mapping can be used without any problem, as we will show later. 

The new method comes from the observation that we can assign any point in an image a material property. For instance, consider mirror reflection; any point on a mirror surface is a reflection of an object. We can simply view that the material property of that particular surface point is the material of the reflected object. Similarly, we can also view the refraction in the same way. The material property can be any shading function. In our case, since we want to obtain a painterly look, we use the simplest barycentric shading, which is given as an interpolation of two colors. In other words, the barycentric material can be represented by two images that define two colors for any given pixel. 

In the simplest barycentric shader, we can use any rendering as the interpolation parameter since rendering provides the amount of light that can reach the surface. The rendering can be computed for any algorithm from photon mapping to radiosity. The effects we can obtain depend only on the power of the rendering algorithm. 

The key idea in this framework is that the final image can be computed in the compositing stage. This is very useful since it is possible to update the results without re-computing rendering again and again.  In the methodology, we provide all the details of this approach.  

\section{Previous Work}

An important direction in the generation of painterly images is the development of painting systems that provide a realistic virtual environment to artists with appropriate brushes and paints by simulating the physical behavior of the actual painting process. A wide variety of tools and systems have been developed for Chinese painting \cite{wu2017chinese}. \cite{ching2002} developed a basic Chinese painting system. 
\cite{xu2003advanced} designed a realistic virtual brush for Chinese painting. \cite{xu2004virtual} developed a virtual hairy brush. \cite{mi2004droplet} developed another virtual brush model to simulate Chinese calligraphy and painting.
\cite{wang2007image} developed a method for the color ink diffusion process common in Chinese painting.
\cite{yin2005hua} implemented an interactive calligraphy and ink-wash painting system. 
\cite{yang2013animating} obtained animating Chinese ink painting by generating reproducible brush strokes.
\cite{huang2019research} developed virtual brush modeling in digital calligraphy and painting.
\cite{yang2020brushwork} obtained Chinese ink painting to animate the brushwork process. 

Another direction is the development of rendering systems to create dynamic Chinese paintings based on 3D scenes with basic local illumination effects \cite{wu2017chinese, xu2009introduction}.  \cite{ching2002} created an animation with some common elements of a Chinese painting. \cite{cheok2007humanistic} created Humanistic Oriental art with automated computer processing and expressive rendering. 
\cite{yuan2007gpu} developed a real-time rendering system to generate a Chinese ink and wash cartoon.
\cite{qi2010research} created Chinese ink animations with a feature-based approach.
\cite{jeong2009oriental} implemented an oriental black-ink rendering system. 
\cite{sun2010multi} developed a multilayer method to render Chinese ink-wash paintings. \cite{li2013rendering, li2020real} developed a few methods for the real-time rendering of 3D animal models in Chinese ink painting style. 
\cite{wang2014stylization} developed a stylization method for Chinese ink painting for 3D trees. 
\cite{gao2022metanalysis} created a large video installation based on the movement of traditional Chinese ink painting. 
\cite{li2023rendering} focused on the rendering and presentation of Chinese landscape paintings and \cite{cheng2023vr} developed VR-based line drawing tools for Chinese painting.
For a detailed overview of most of this work based on conversations with specialists, see \cite{zhang2023computational}.

Our work involving the inclusion of global illumination effects in Chinese painting follows the work of \cite{liu2015}. The main issue with this approach is that it requires a separate composition program, such as Nuke \cite{liu2015a}. Our goal in this work is to further simplify the process. Our approach is image-based and conceptually uses a 2.5D pipeline \cite{akleman2022dynamic,akleman2023web}.

\section{Methodology}

The process of obtaining dynamic still-life painting with global effects usually consists of four steps \cite{liu2015,justice2018,yan2015,subramanian2020painterly,ross2021Georgia}: (1) Modeling proxy geometry and reflective properties, (2) Creation of control paintings, (3) Projecting the control paintings to proxy geometry, (4) rendering images with a renderer that supports global illumination such as Renderman \cite{christensen2018renderman,apodaca1999advanced} or Arnold \cite{georgiev2018arnold,kulla2018sony}. 

The main problem in this process is the fourth step. Rendering software such as Renderman and Arnold is mainly developed to support physically accurate renderings. To use them in an expressive setting, users need to know how to program them in shaders. However, this is possible in big companies, for individual users that are interested in obtaining expressive styles, there is not much option. 

In this paper, we simplify the fourth step by splitting it into two steps, a rendering step followed by a compositing step. An important part of our approach is that the rendering step does not require any programming that is tailored to obtain a specific style. As we have shown later, any rendering (even physically accurate ones) can be used to obtain expressive styles. The compositing step in our approach is essentially related to 2.5D pipelines that use only images \cite{akleman2022dynamic,akleman2023web}.

As a result of splitting the rendering step, i.e. the fourth step, into two, we define a new process that consists of five steps. Each of these five steps is simple and intuitive. Any 3D artist with basic training in 3D modeling and animation can potentially produce such dynamic still-life paintings using this process. The following lists all the steps of the process.  

\begin{enumerate}
\item \textbf{Modeling proxy geometry:} This step involves designing the appropriate proxy geometry and reflective properties such that the shape is correctly registered from a particular viewpoint. 
\item \textbf{Modeling Control Textures:} This step is a painting step. Using one of the painting systems, users need to create a set of control images that will be used as textures. 
\item \textbf{Projecting Control Textures:} This process is the projecting control paintings as textures on proxy geometry. Since proxy geometry is created by correctly registering the shapes from a particular point of view, this texture mapping process is also supposed to be straightforward. 
\item \textbf{Rendering:} This step involves rendering two control texture animations without light and one illuminated animation using only one material per object. This step can be implemented in any modeling and rendering software, such as Maya or Blender, and it does not require a more sophisticated renderer such as Renderman or Arnold. 
\item \textbf{Composting:}  This step is simple interpolation of the two control texture animations using illuminated animation as weights of the interpolation. This step can be implemented in animation software such as AfterEffect. 
\end{enumerate}

 \begin{figure*}[htb]
  \centering
        \begin{subfigure}[t]{0.49\textwidth}
        \includegraphics[width=1.0\textwidth]{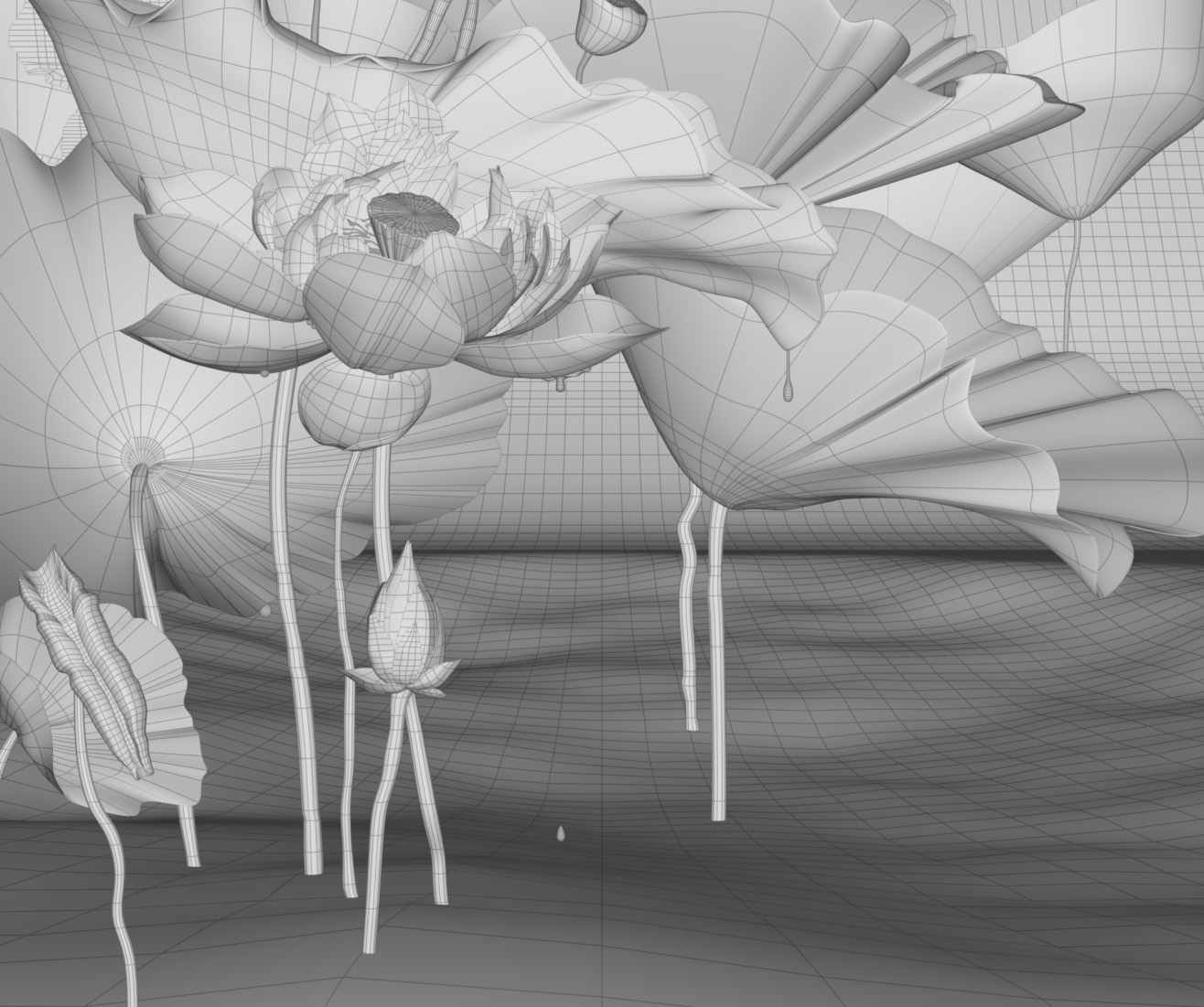}
        \caption{Proxy Geometry Front View.}
        \label{fig_proxyfrontview}
    \end{subfigure}
        \hfill
         \begin{subfigure}[t]{0.49\textwidth}
        \includegraphics[width=1.0\textwidth]{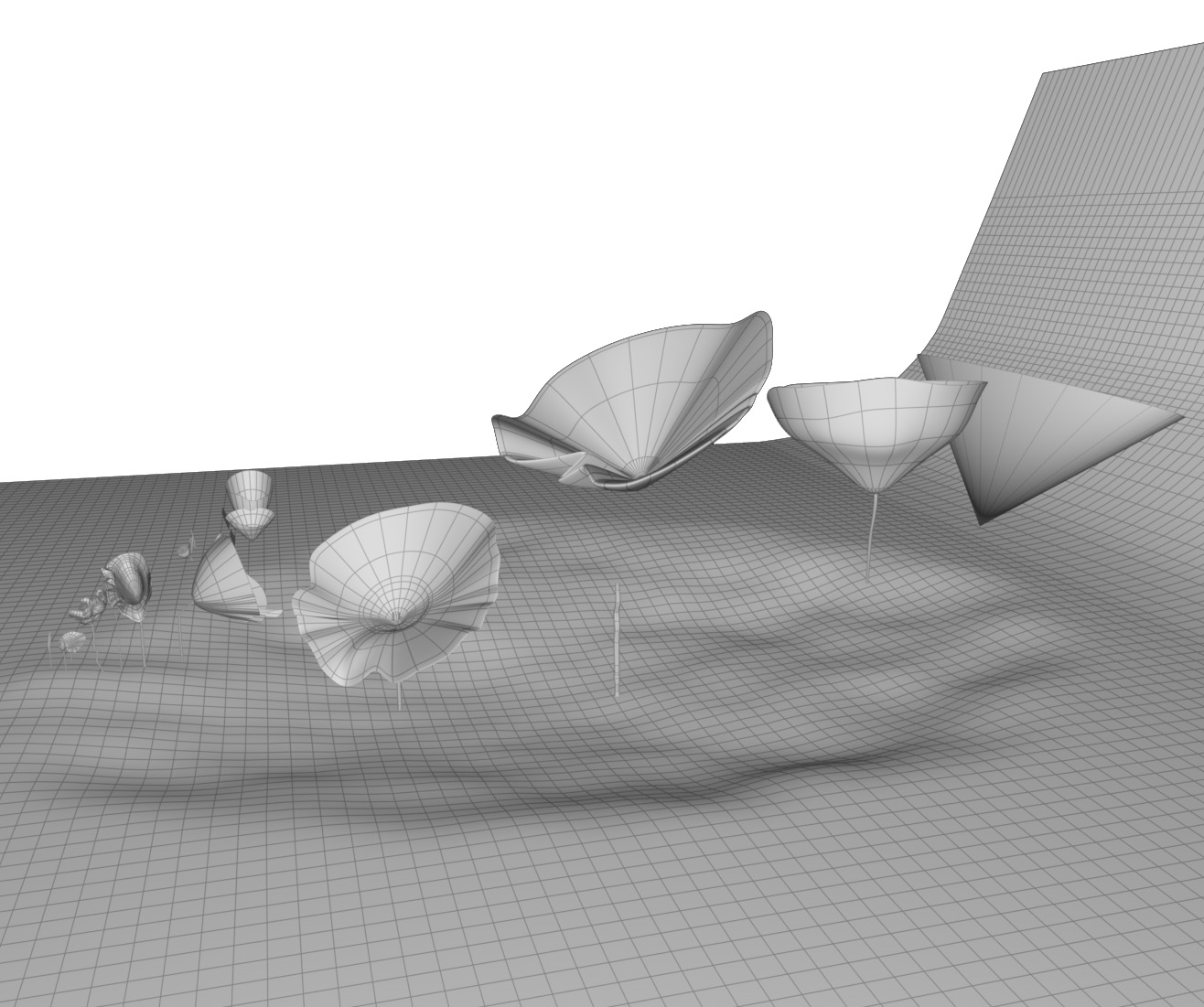}
        \caption{Proxy Geometry Side View.}
        \label{fig_proxysideview}
    \end{subfigure}
    \caption{\it Anamorphic Bas-Relief type proxy geometry.}
\label{fig_proxy}
\end{figure*}

\subsection{Modeling Proxy Geometry}

The first step in our process is modeling proxy geometry and reflective properties. The key idea in this step is to have correct registrations. In other words, the boundaries of the 3D models match the silhouette edges of the objects in the original painting so that the projections of the diffuse and shadow images onto the models will line up precisely (see Figures~\ref{fig_proxyfrontview}, and~\ref{fig_proxysideview}). 

As long as the boundaries of the models match those of the control images, the rest of the geometry can be a quick approximation of a 3D form. These shapes do not need to be perfectly accurate. Note that in Figure~\ref{fig_proxyfrontview} the geometry looks good from the front view, which we will render from. However, it is an approximation from the side view, as shown in Figure~\ref{fig_proxysideview}. We also assign the values of the reflection coefficient $k_s$ and the index of the refraction coefficient $\eta$ to each object to control the strength of the reflections.  

     \begin{figure*}[htb]
  \centering
        \begin{subfigure}[t]{0.61\textwidth}
        \includegraphics[width=1.00\textwidth]{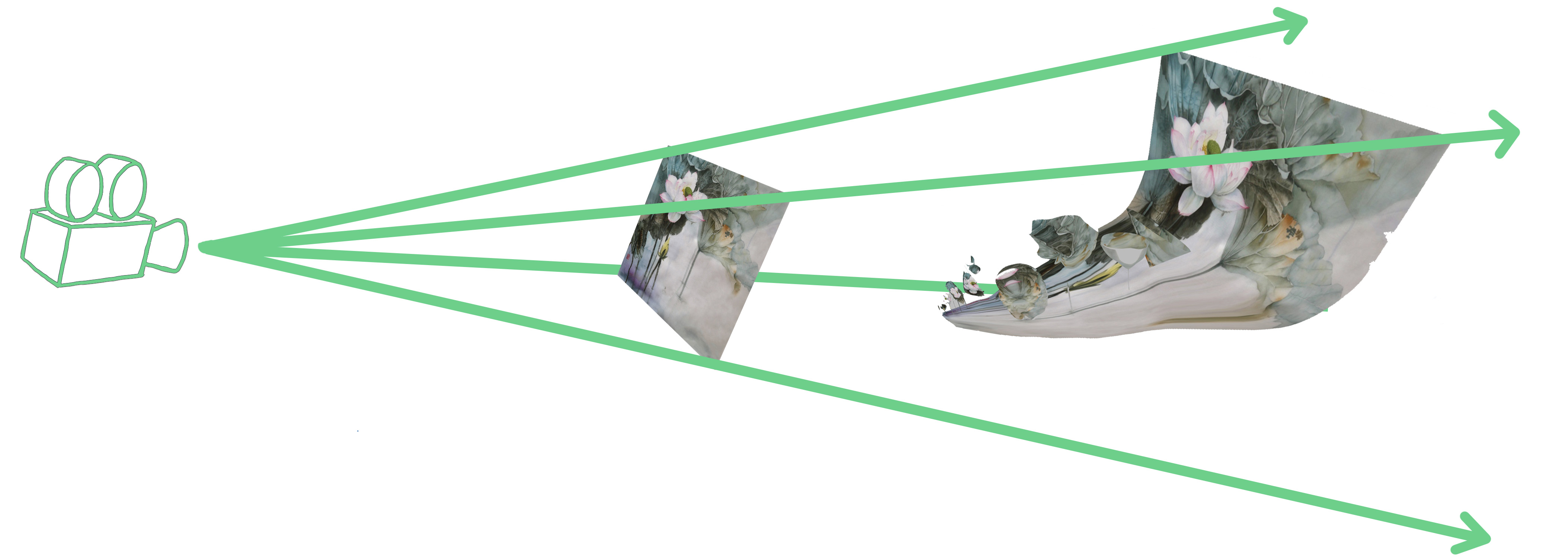}
        \caption{ Projection of textures to proxy geometry.}
        \label{fig_projection}
    \end{subfigure}
     \hfill
    \begin{subfigure}[t]{0.38\textwidth}
        \includegraphics[width=1.00\textwidth]{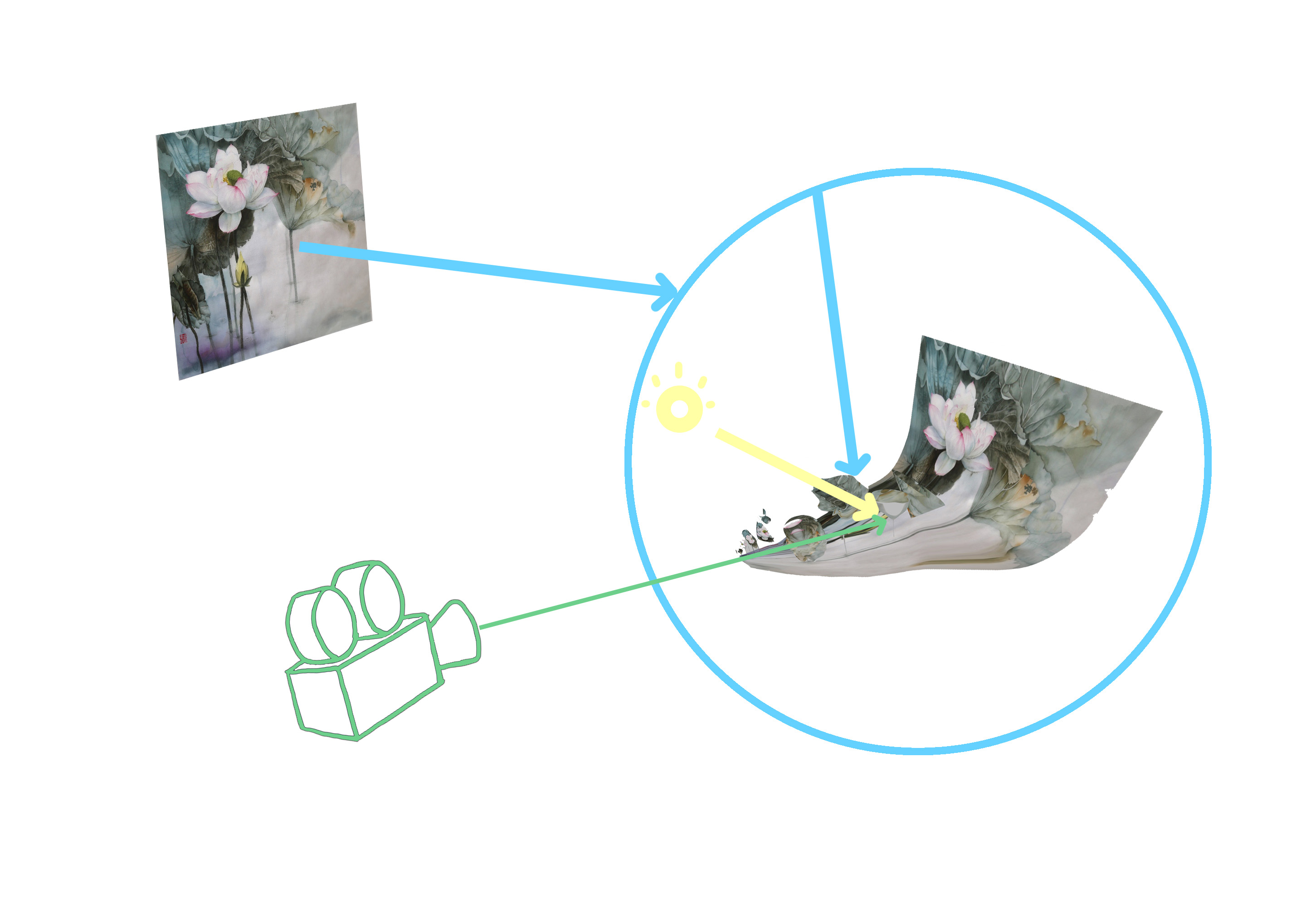}
        \caption{  Reflections are computed using textures.}
        \label{fig_reflection visualization}
    \end{subfigure}
\caption{\it Projection of textures to proxy geometry.}
\label{fig_proxy2}
\end{figure*}

\subsection{Creation of control Paintings}

The second step of the process is to create control paintings.  If there is only one key (strong) light in the scene \cite{akleman2016}, we just need two types of control paintings, one type for shadows and another type for fully illuminated diffuse. These two types of texture images can be created using any digital painting software. For fully illuminated diffuse painting, the goal is to make every part of the image look as though it were illuminated by the light source. To accomplish this, we can paint over or color correct areas that were painted to be in shadow or had a specular highlight, reflection, or caustic. For the shadow type of images, we need to do the opposite but still eliminate all the global illumination effects. 

\subsection{Projections of control Paintings}

Projecting these paintings onto proxy geometry is simple. Since we have created proxy geometry that is mainly correct from one point of view and the boundaries of the proxy geometry match with these paintings, we can simply map these paintings onto the proxy geometry using a simple camera projection, as shown in Figure~\ref{fig_projection}. Both diffuse and shadow-type images will have separate projections.

\subsection{Rendering}

\begin{figure}[htb!]
  \centering
        \begin{subfigure}[t]{0.48\textwidth}
        \includegraphics[width=1.0\textwidth]{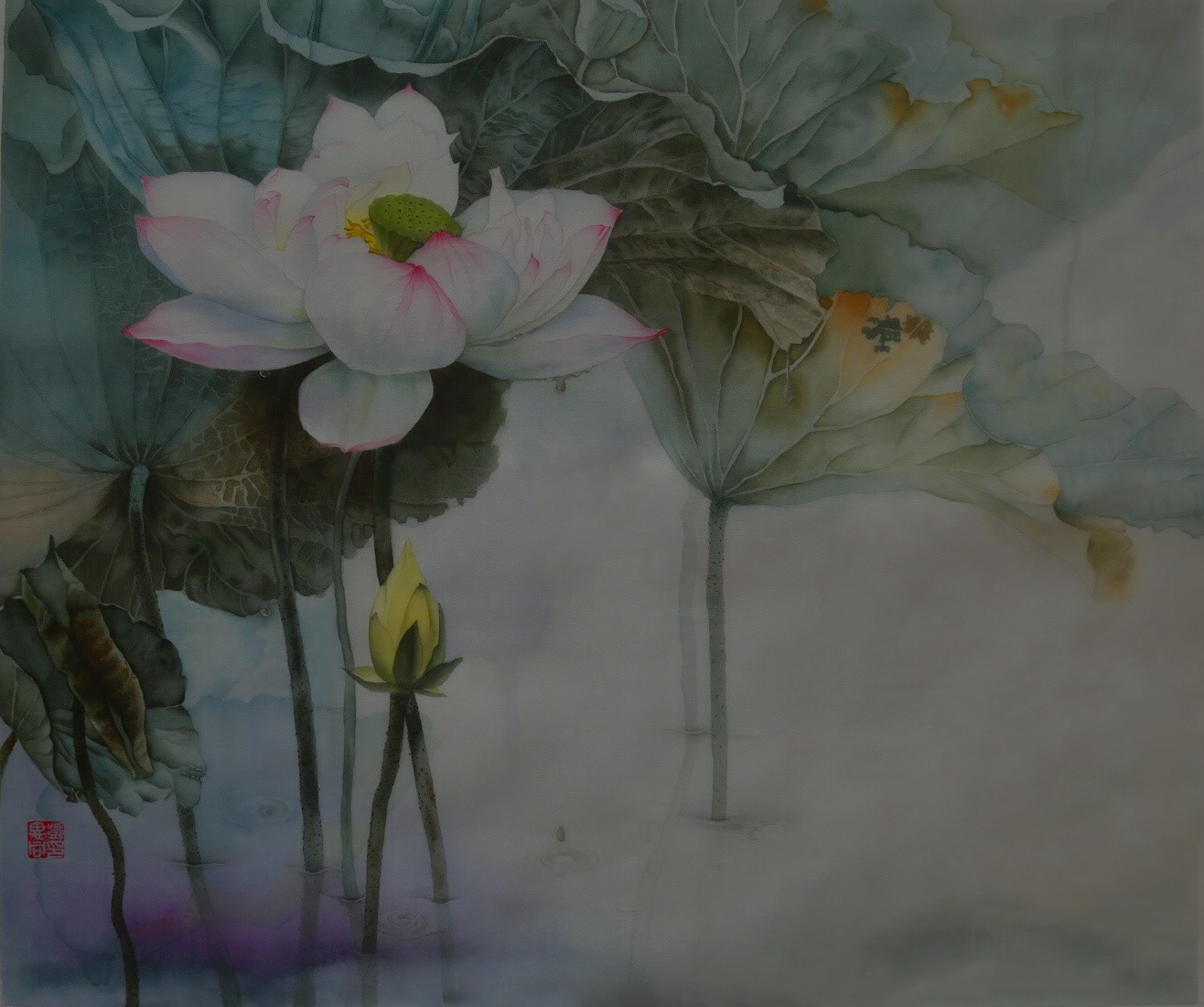}
        \caption{A frame of the shadow animation $T_{0}(u,v,t)$.}
        \label{fig_t0}
    \end{subfigure}
    \hfill
            \begin{subfigure}[t]{0.48\textwidth}
        \includegraphics[width=1.0\textwidth]{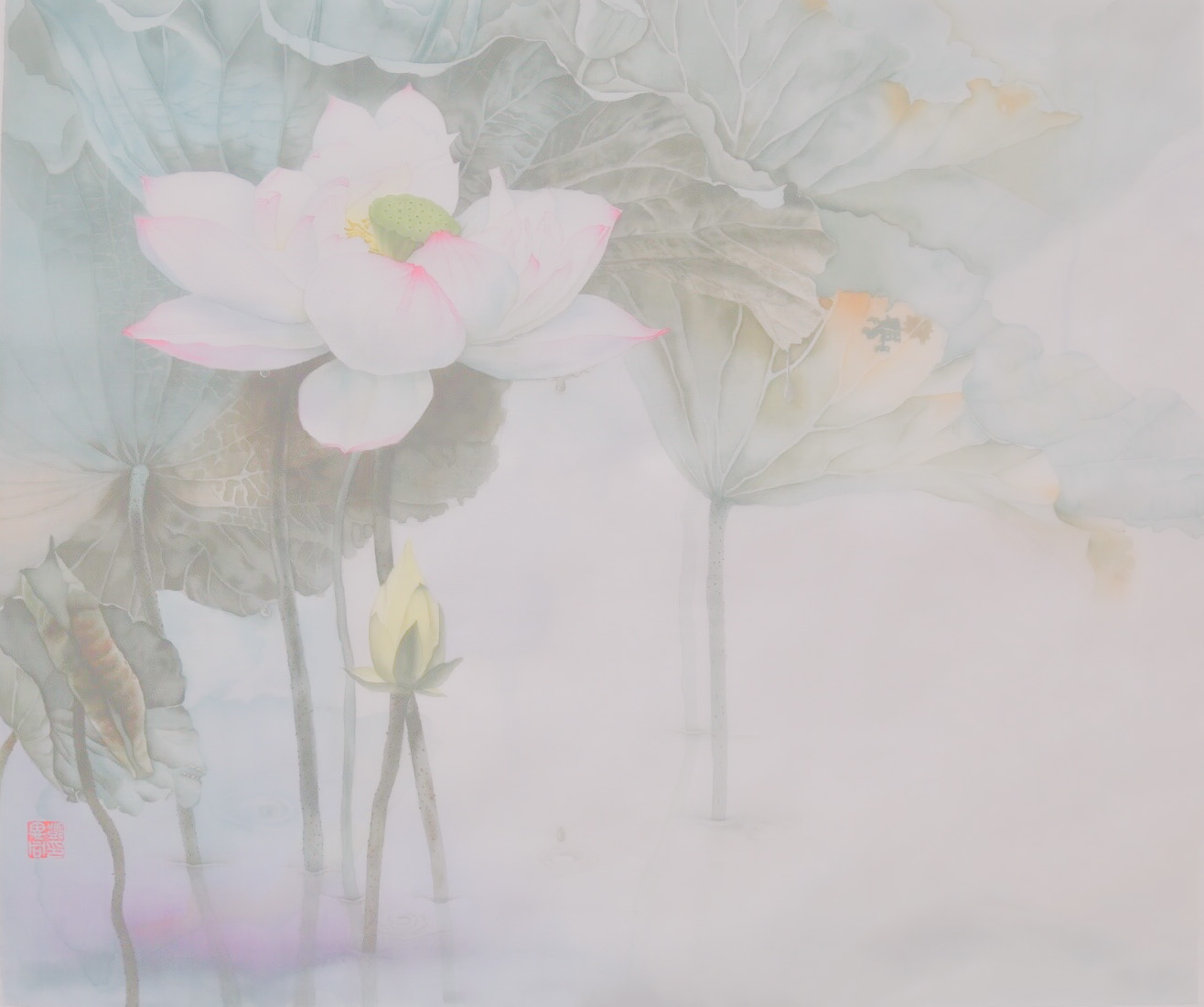}
        \caption{A frame of $T_{1}(u,v,t)$.}
        \label{fig_t1}
    \end{subfigure}
    \hfill
        \begin{subfigure}[t]{0.48\textwidth}
        \includegraphics[width=1.00\textwidth]{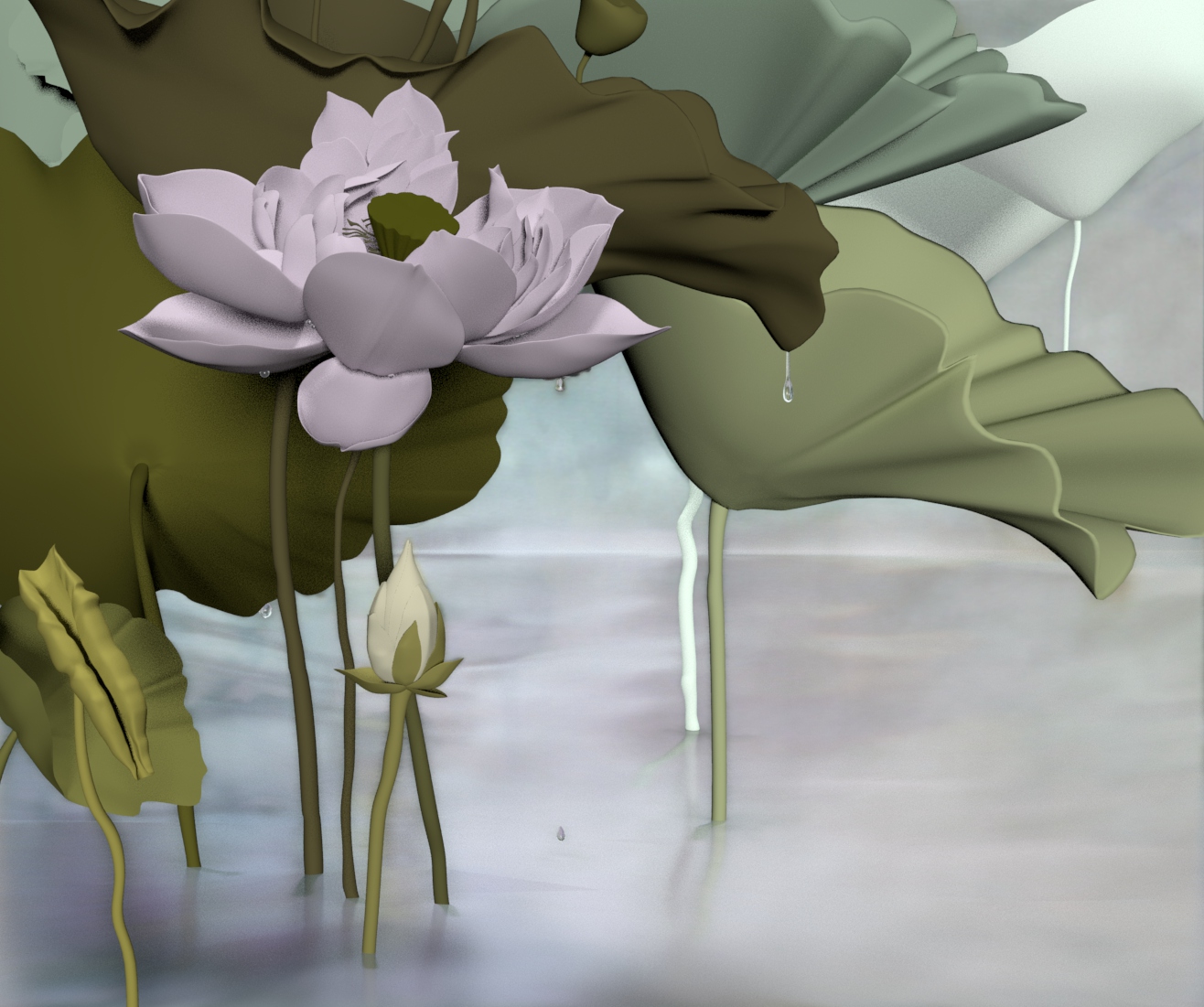}
        \caption{ A frame of weight animation,  $W(u,v,t)$.}
        \label{fig_w}
    \end{subfigure}
        \hfill
        \begin{subfigure}[t]{0.48\textwidth}
        \includegraphics[width=1.00\textwidth]{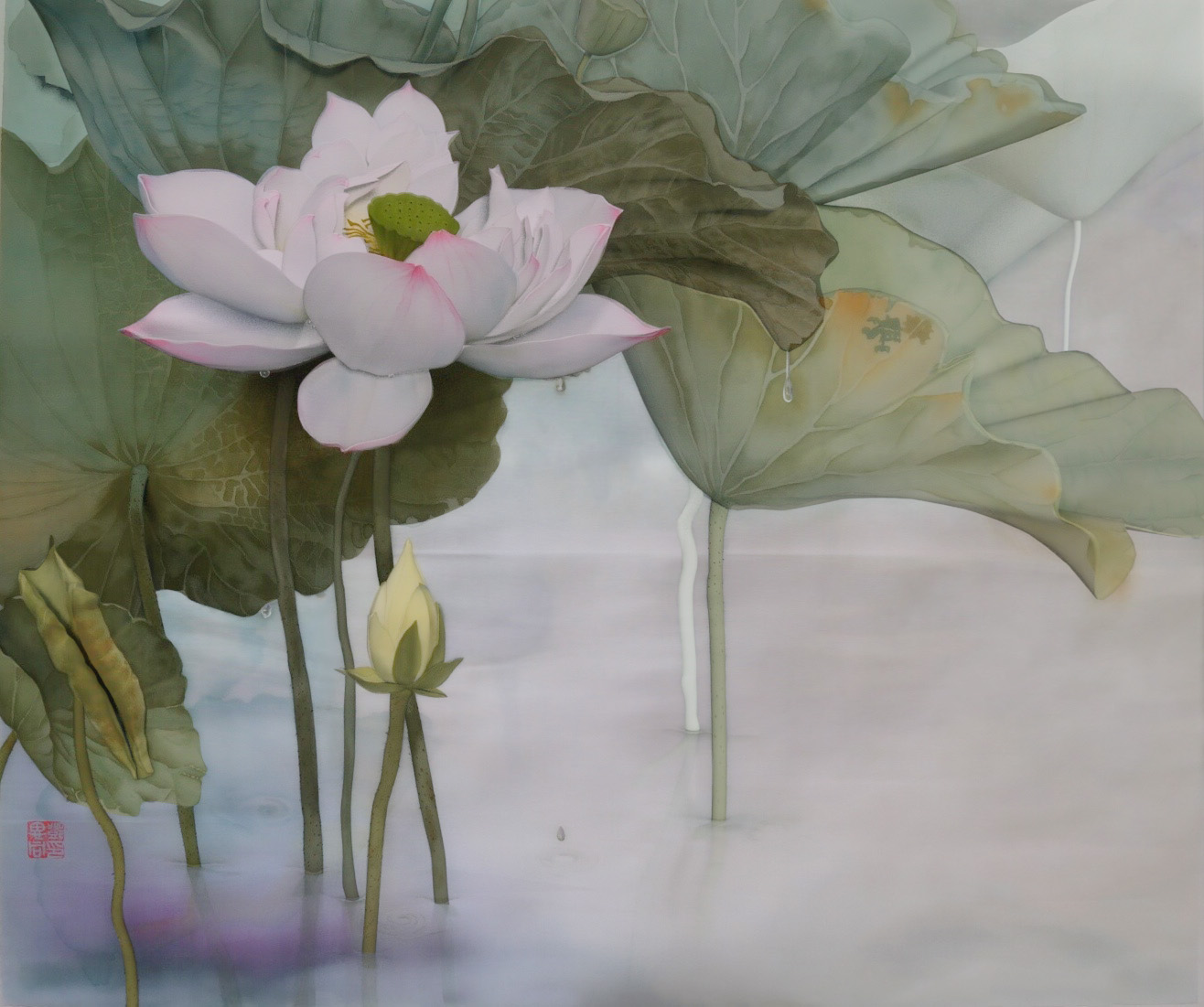}
        \caption{ A frame of composited animation,  $C(u,v,t) = T_{0}(u,v,t) W(u,v,t) + T_{1}(u,v,t) (\mathbf{I} - W(u,v,t))$.}
        \label{fig_c}
    \end{subfigure}
\caption{\it Two texture images $T_{0}$ and $T_{1}$ and front and side views of the proxy geometry that is used to compute illumination and shading.}
\label{fig_textures}
\end{figure}

The next step of the process is to render three animations using moving key light and moving objects. Let $T_{0}(u,v,t)$, $T_{1}(u,v,t)$, and $W(u,v,t)$ denote the three animations, where $(u,v)$ denote pixel coordinates and $t$ denote time. The shadow animation, $T_{0}(u,v,t)$, is rendered using only shadow-type textures without any illumination, as shown in Figure~\ref{fig_t0}. Similarly, fully illuminated animation, $T_{1}(u,v,t)$, is rendered using only fully illuminated diffuse texture types without any illumination, as shown in Figure~\ref{fig_t1}. Note that this rendering process also includes global illumination effects such as reflections and refractions such that they are embedded in $T_{0}(u,v,t)$ and $T_{1}(u,v,t)$. The last animation, $W(u,v,t)$, is calculated using moving light, as shown in Figure~\ref{fig_w}. We assign a single material to each object. We do not use any of the textures to produce this animation, which we call weight animation, since it is used only as a Barycentric weight.

\subsection{Compositing}

The last step of the process is to compose to obtain the final animation $C(u,v,t) $. The composition is an interpolation of $T_{0}(u,v,t)$ and $T_{1}(u,v,t)$ using $W(u,v,t)$ as a Barycentric weight \cite{akleman2016})  by following equation: 
$$
C(u,v,t) = T_{1}(u,v,t) W(u,v,t) + T_{0}(u,v,t) (\mathbf{I} - W(u,v,t)).
$$
where $\mathbf{I}$ is a white image and multiplication is simply the multiplication of each color channel for every pixel. Figures~\ref{fig_t0},~\ref{fig_t1}, and~\ref{fig_w} show $T_{0}(u,v,t)$, $T_{1}(u,v,t)$, and $W(u,v,t)$, respectively. It is interesting to note that $C(u,v,t)$ shown in Figure~\ref{fig_c} looks much richer than $T_{0}(u,v,t)$, $T_{1}(u,v,t)$ and $W(u,v,t)$. More results corresponding to different values of $t$ are shown in Figures~\ref{fig_teaser}, and ~\ref{fig_teaser2} as frames of the animated painting. 

\section{Discussion}

\begin{figure}[htb!]
  \centering
        \includegraphics[width=1.0\textwidth]{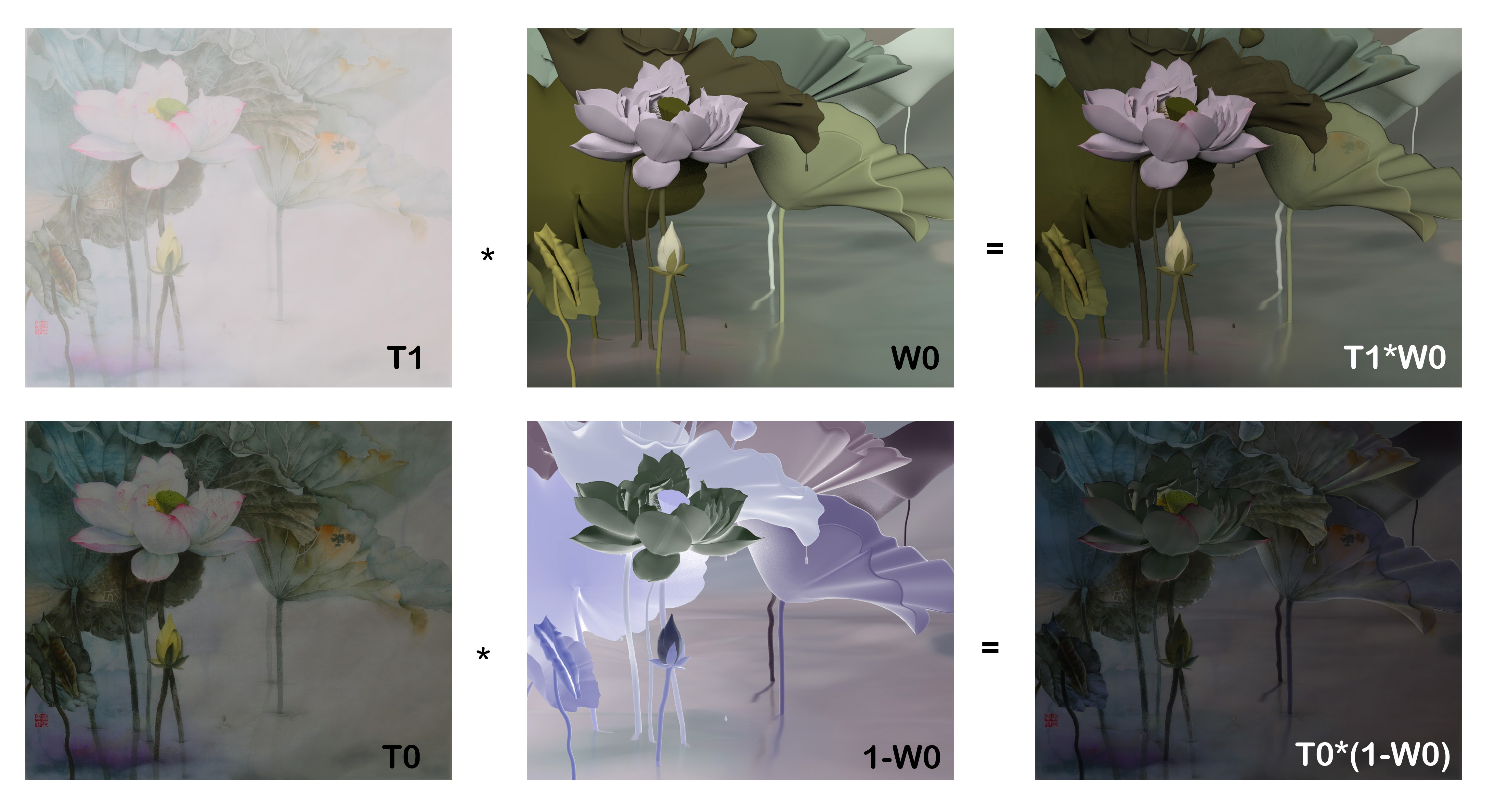}
        \includegraphics[width=1.0\textwidth]{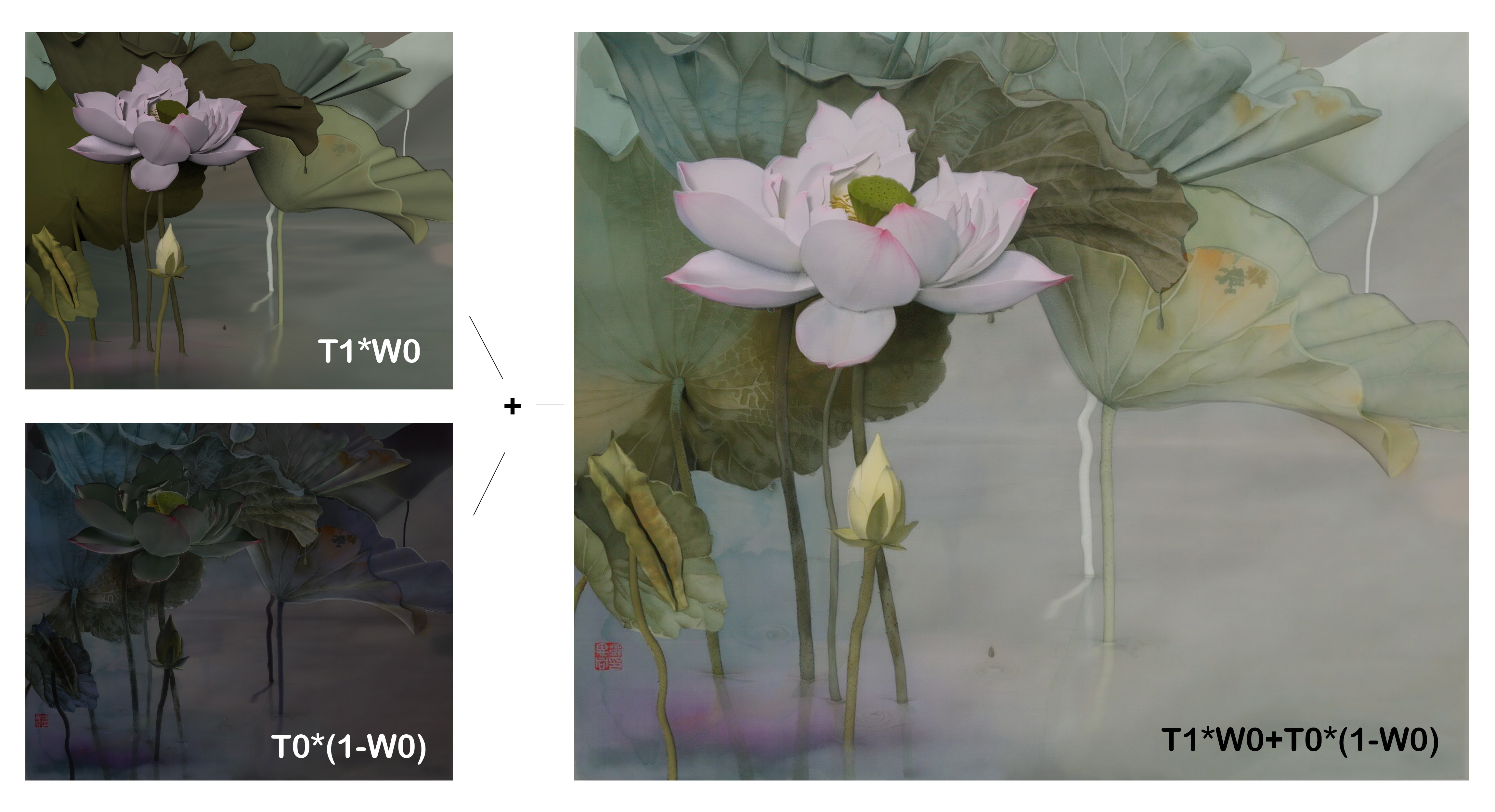}
\caption{Compositing operation with a manipulated $W(u,v,t)$ image.}
\label{fig_compositing0}
\end{figure}

\begin{figure}[htb!]
  \centering
 \includegraphics[width=1.0\textwidth]{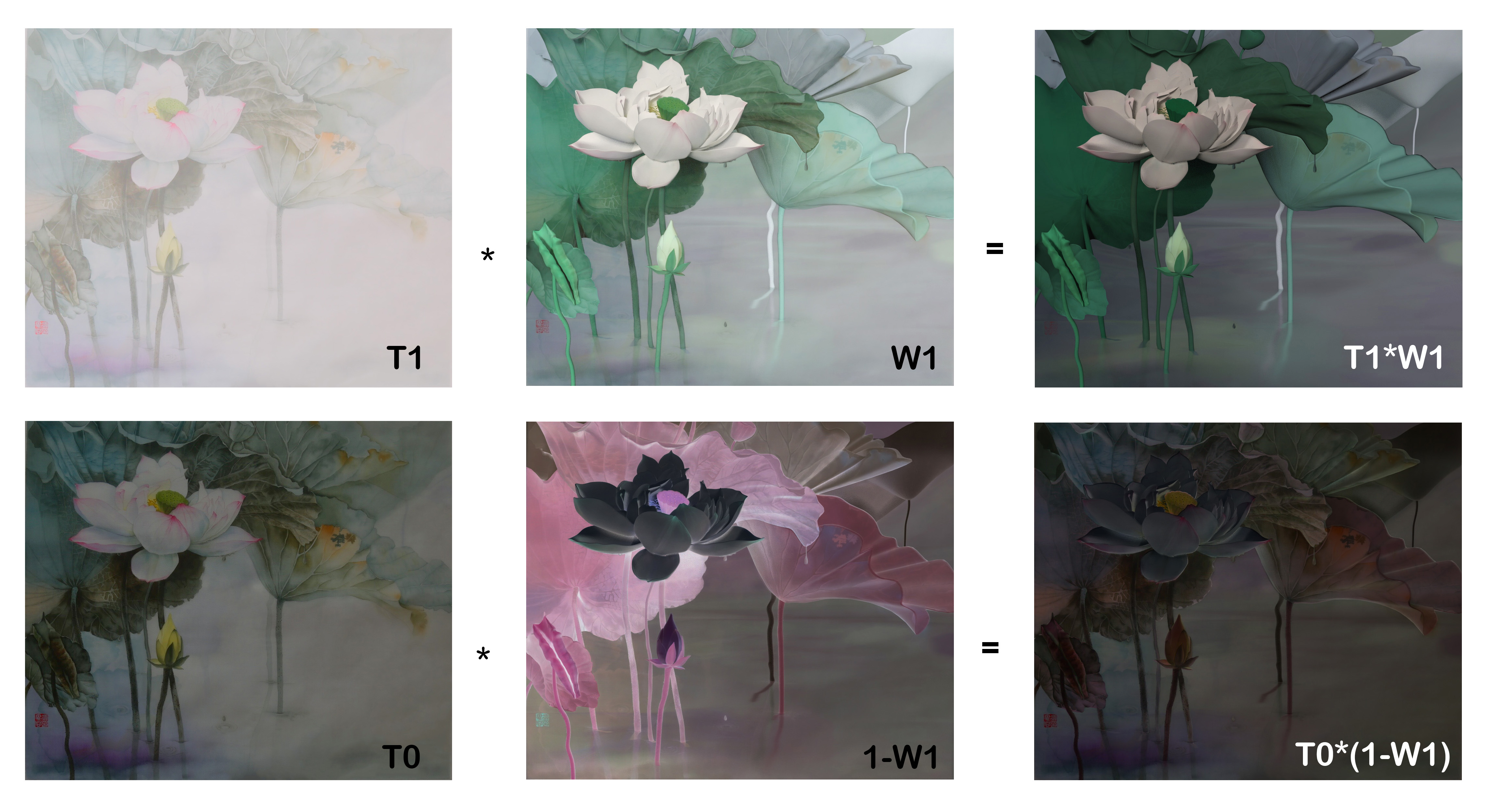}
        \includegraphics[width=1.0\textwidth]{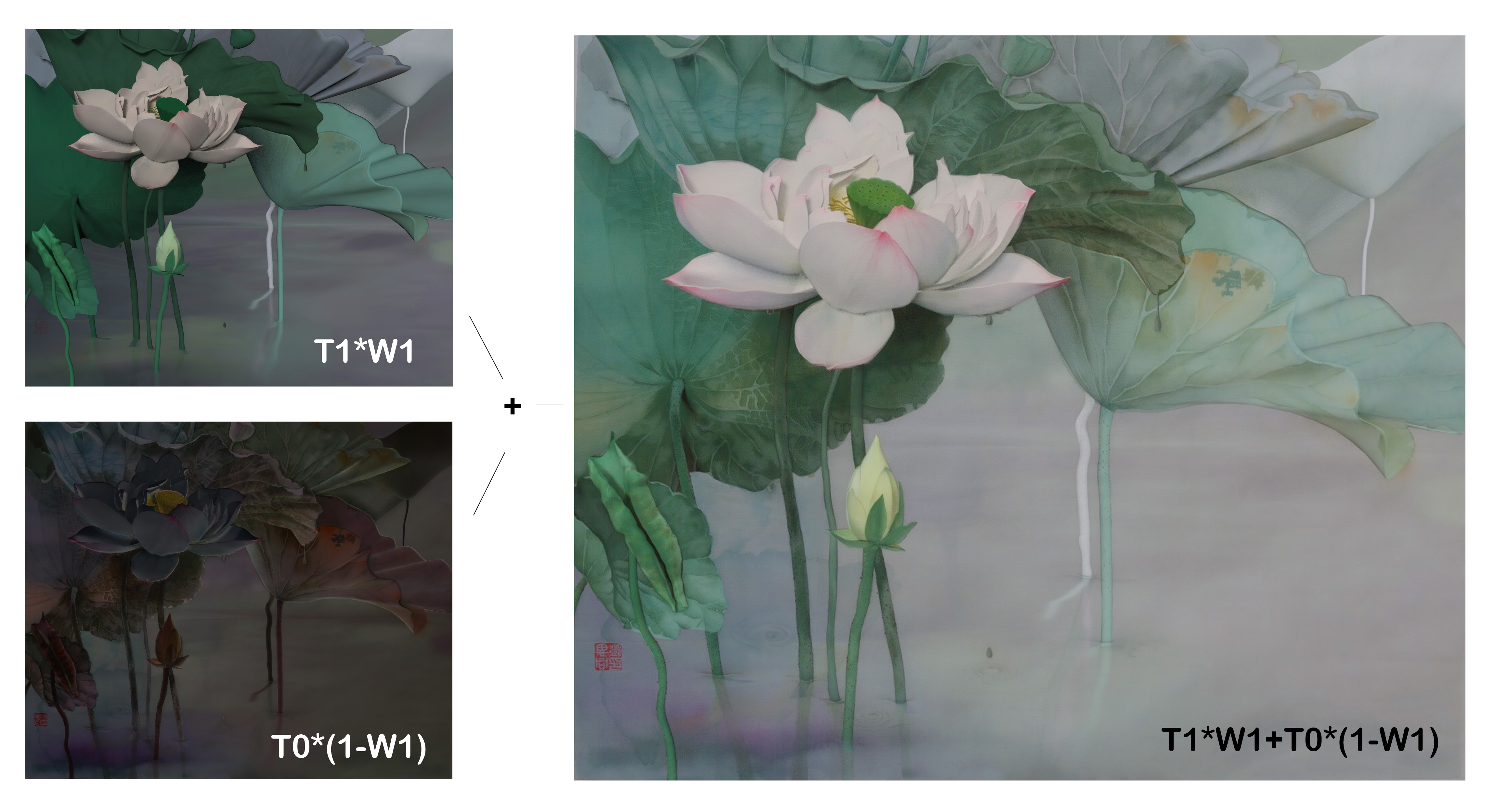}
\caption{Compositing operation with a manipulated $W(u,v,t)$ image. In this case, we changed the hue of the $W(u,v,t)$ image to obtain a more greenish look.}
\label{fig_compositing1}
\end{figure}

\begin{figure}[htb!]
  \centering       \includegraphics[width=1.0\textwidth]{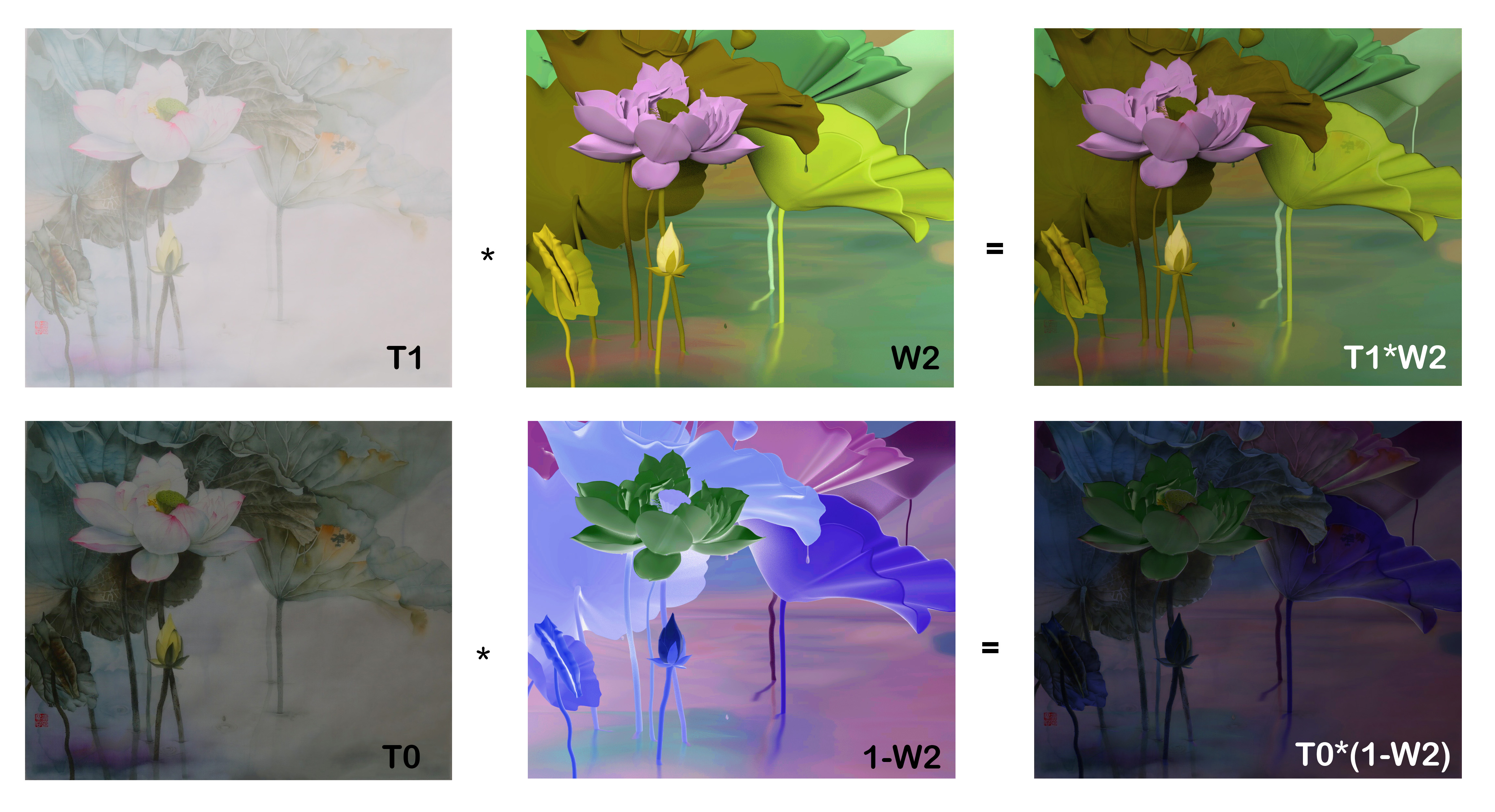}       \includegraphics[width=1.0\textwidth]{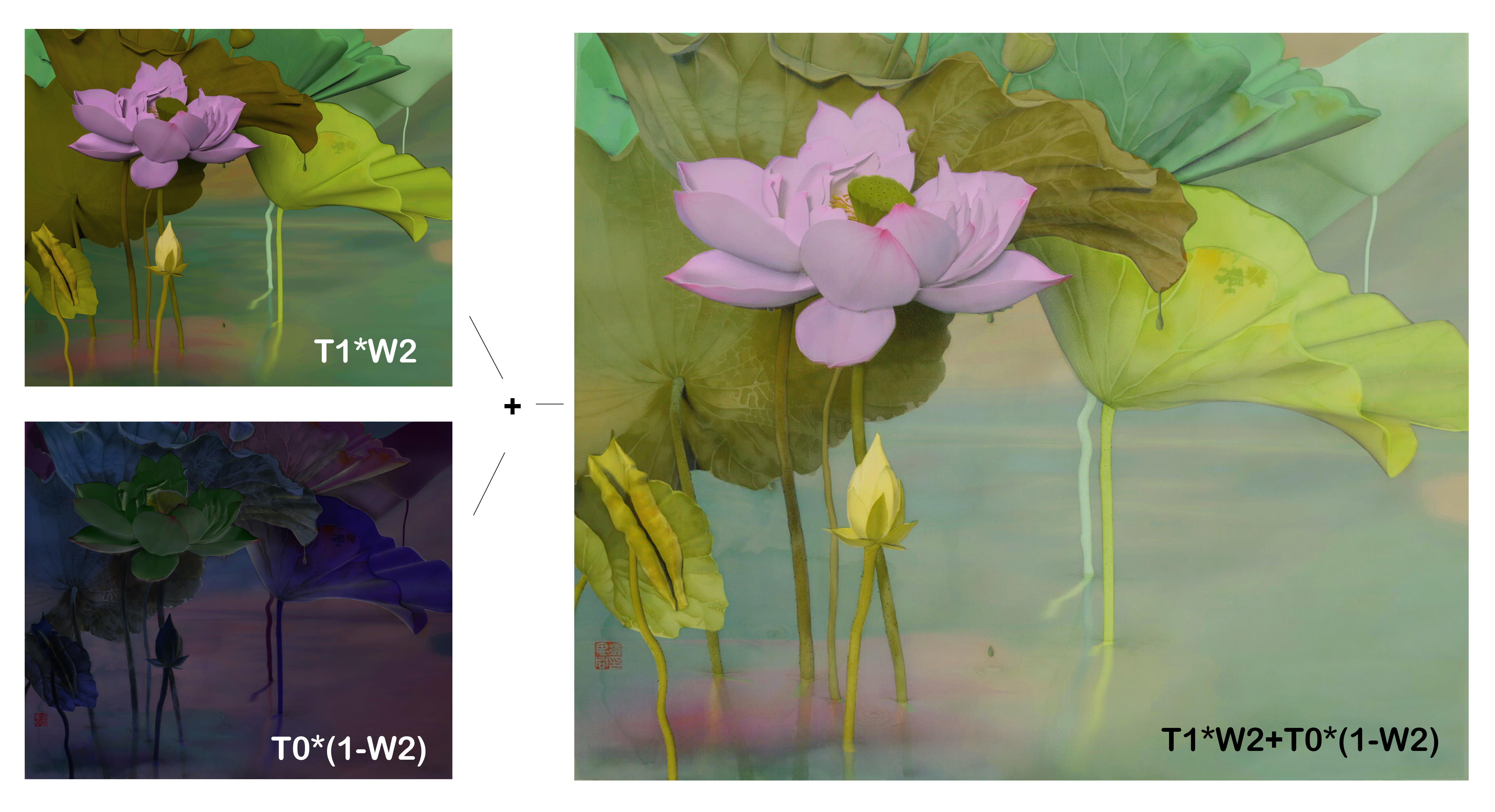}
\caption{Compositing operation with a manipulated $W(u,v,t)$ image. In this case, we changed both the hue and saturation of the image $W(u,v,t)$ to obtain a very saturated and colorful result. }
\label{fig_compositing2}
\end{figure}

\begin{figure}[htb!]
  \centering       \includegraphics[width=1.0\textwidth]{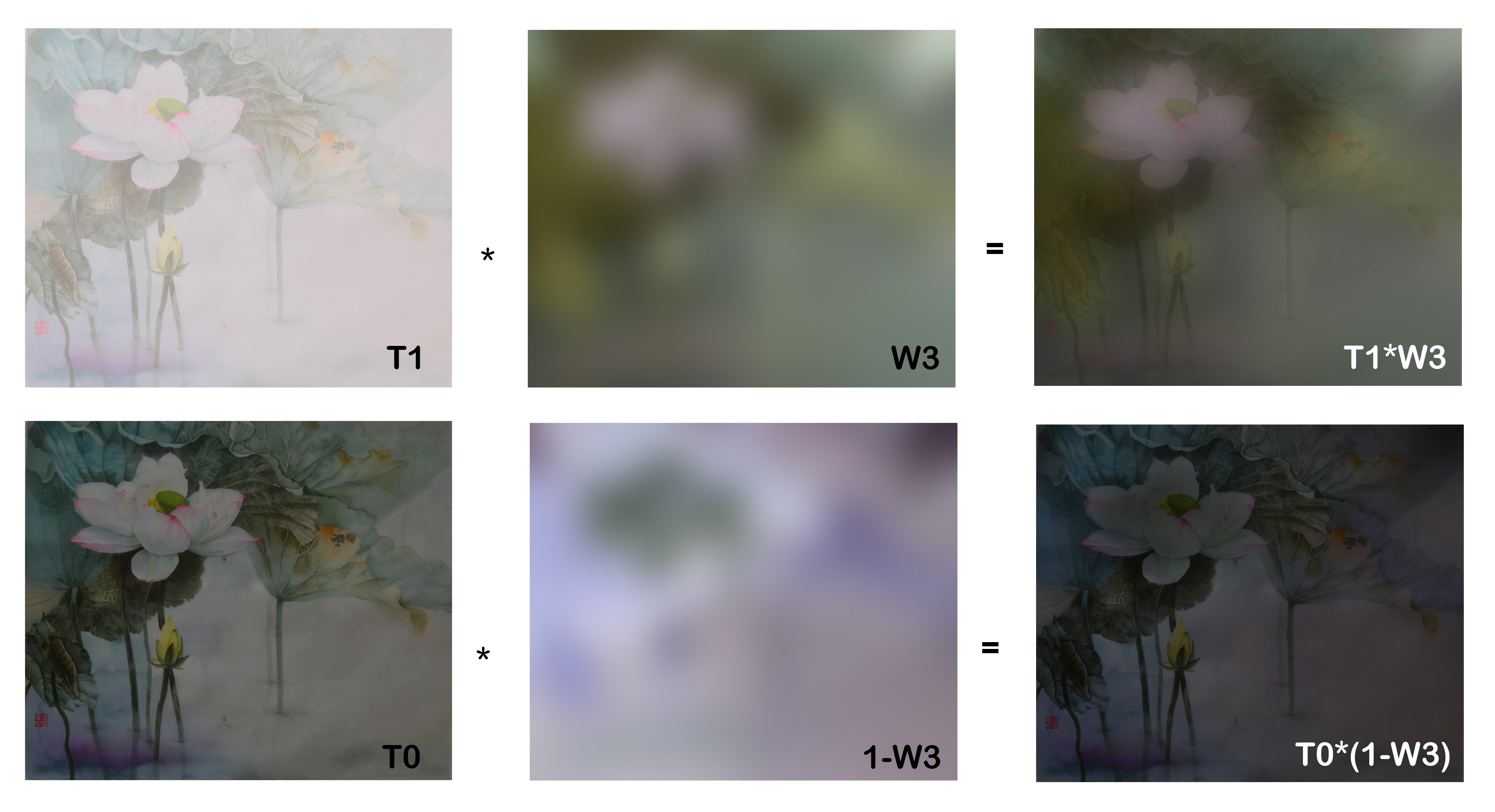}       \includegraphics[width=1.0\textwidth]{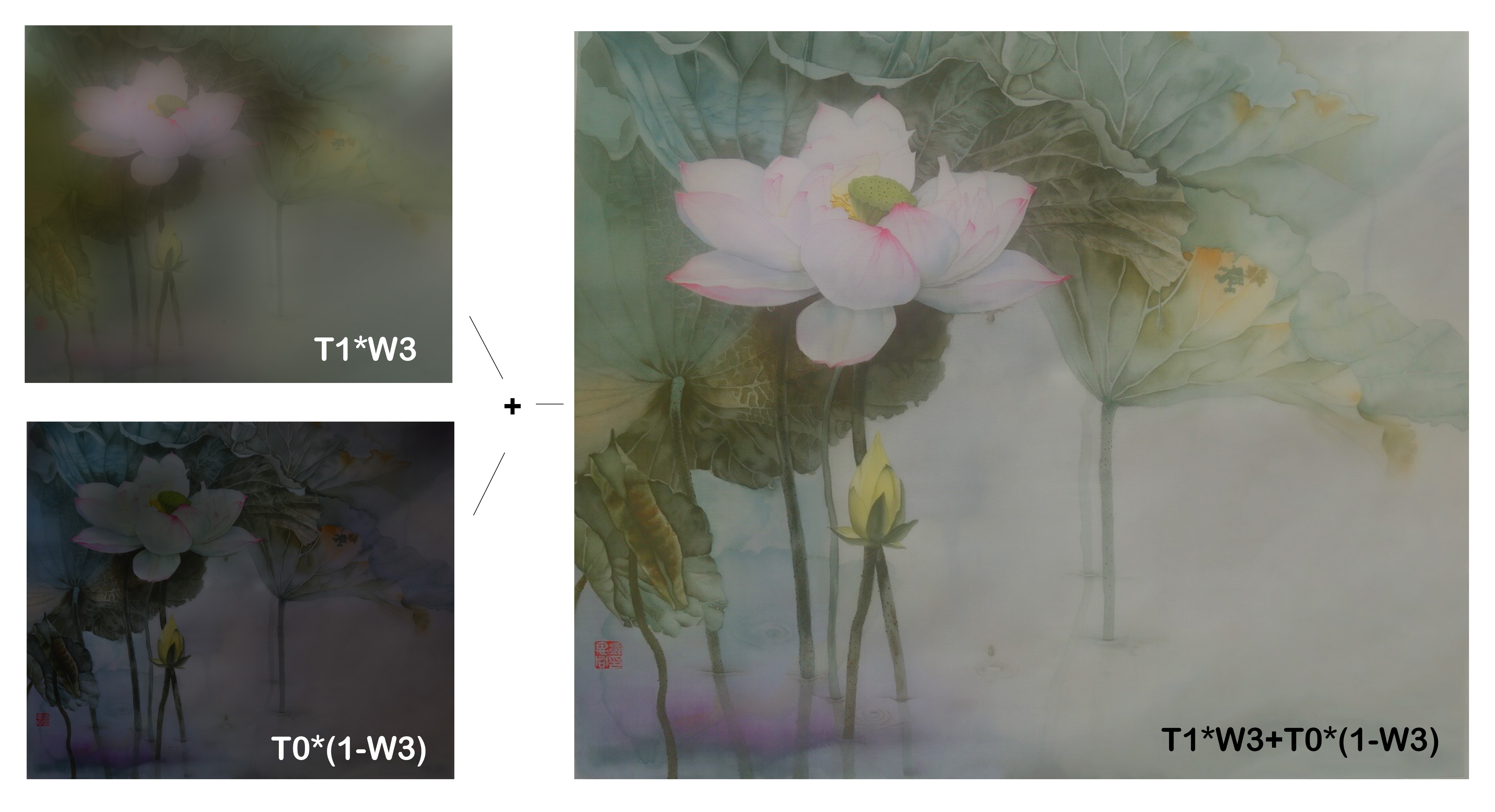}
\caption{Compositing operation with a manipulated $W(u,v,t)$ image. In this case, we blurred $W(u,v,t)$ to obtain show that the result is robust. Note that the term that corresponds to the classical (non-barycentric) diffuse formula $T_1W_3$ is just a blurred image. On the other hand, the barycentric formula still provides clean results. }
\label{fig_compositing3}
\end{figure}

\begin{figure}[htb!]
  \centering       \includegraphics[width=1.0\textwidth]{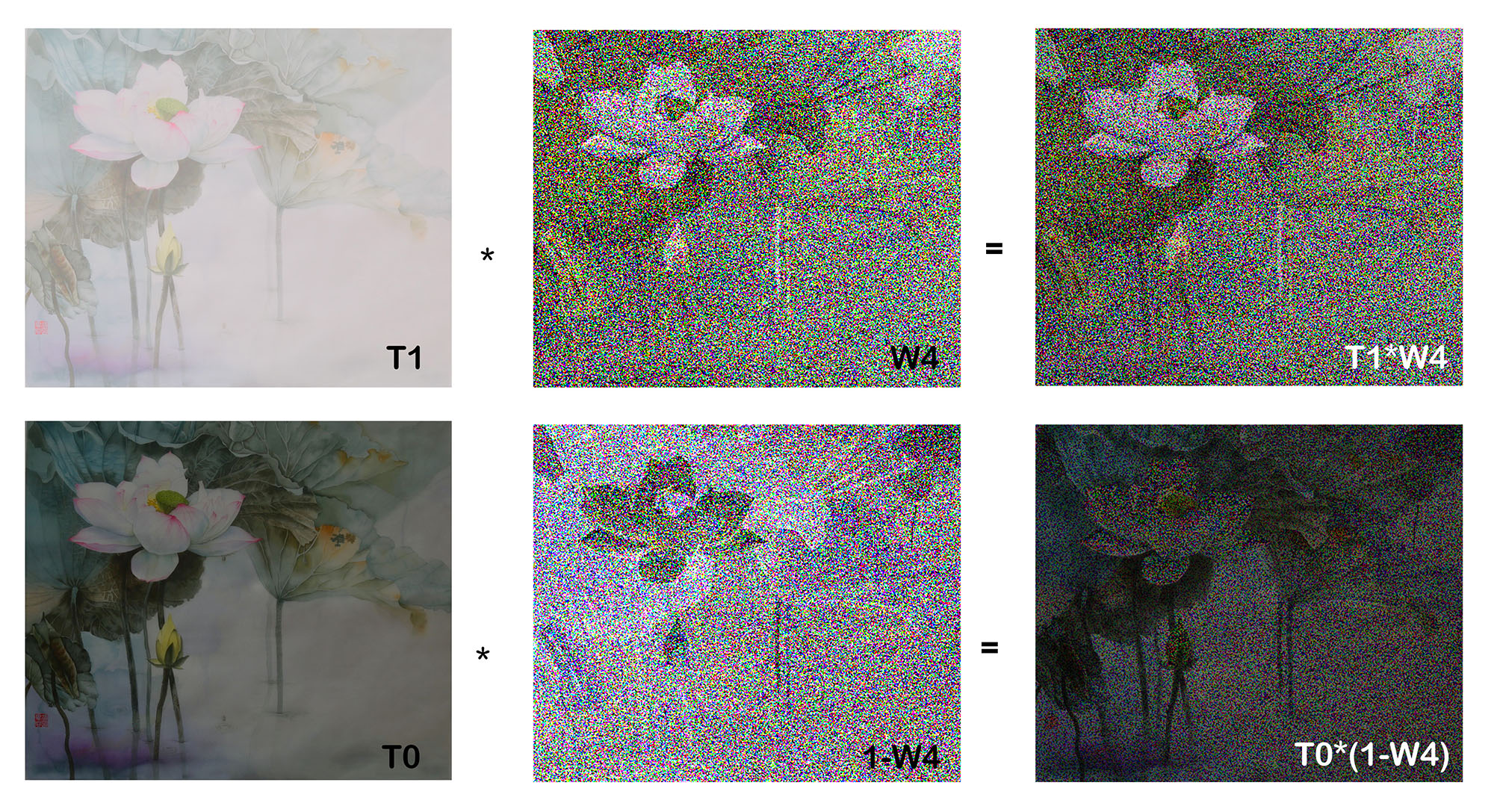}       \includegraphics[width=1.0\textwidth]{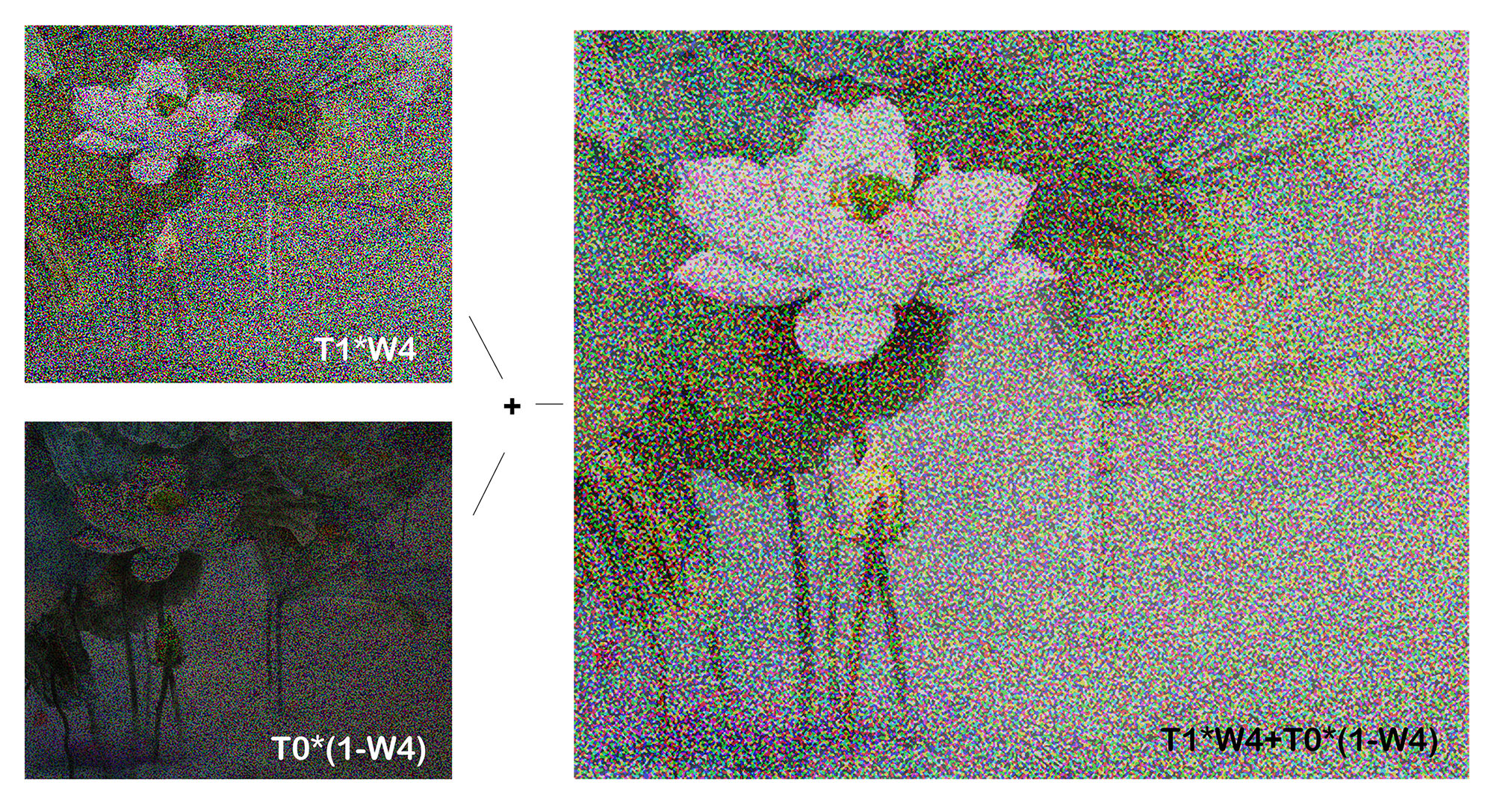}
\caption{Compositing operation with a manipulated $W(u,v,t)$ image. In this case, we dithered $W(u,v,t)$ to obtain a pointillist style. }
\label{fig_compositing4}
\end{figure}

An important property of this approach is that we can manipulate the final result simply by manipulating $W(u,v,t)$. Figures~\ref{fig_compositing0},~\ref{fig_compositing1},~\ref{fig_compositing2},~\ref{fig_compositing3}, and~\ref{fig_compositing4} show the effect of image manipulation $W(u,v,t)$. An important property of this process regardless of the manipulation is that most of the resulting images still appear to be of the same style. The only exception is Figure~\ref{fig_compositing4}, in which the resulting painting appears to be pointillist style \cite{yang2008realization,wu2013generating}. 

Let us now examine the manipulations closely. Figure~\ref{fig_compositing0} shows the original rendering. In Figure~\ref{fig_compositing1}, we changed the hue of the $W(u,v,t)$ image to obtain a more greenish look. In Figure~\ref{fig_compositing2}, we changed both the hue and saturation of the image $W(u,v,t)$ to obtain a very saturated and colorful result. In Figure~\ref{fig_compositing3}, we blurred $W(u,v,t)$ to show that the result is robust.  In Figure~\ref{fig_compositing4}, we dithered $W(u,v,t)$ to obtain a pointillist style. 

We want to point out that the main reason behind the robustness behind this approach comes from the barycentric formula. The classical formula is a linear formula and is in the form of $C(u,v,t) = T_{1}(u,v,t) W(u,v,t)
$. Note that this corresponds to choosing $T_{0}(u,v,t) =0$. Note that the additional term makes the difference. 

\section{Conclusion and Future Work}

In this work, we present an approach to turn Chinese still life paintings with global illumination effects into dynamic paintings with moving lights. We are able to preserve the original look and feel of still-life paintings with moving lights and objects. Our method is very simple and is formulated as a composite of two animated texture images using an animated rendering. The compositing processes are implemented directly over two animations. This approach can be used to turn almost any still-life painting into a dynamic painting.

\bibliographystyle{ACM-Reference-Format}
\bibliography{references,refthesis} 

\end{document}